# A phylogeny of birds based on over 1,500 loci collected by target enrichment and high-throughput sequencing

John E. McCormack,*[1] Michael G. Harvey,[1,2] Brant C. Faircloth,[3] Nicholas G. Crawford,[4] Travis C. Glenn[5], & Robb T. Brumfield[1,2]

[1] *Museum of Natural Science, Louisiana State University, Baton Rouge, LA 70803, USA*

[2] *Department of Biological Sciences, Louisiana State University, Baton Rouge, LA 70803, USA*

[3] *Department of Ecology and Evolutionary Biology, University of California, Los Angeles, CA 90095, USA*

[4] *Department of Biology, Boston University, Boston, MA 02215, USA*

[5] *Department of Environmental Health Science, University of Georgia, Athens, GA 30602, USA*

*Corresponding author: John McCormack, Moore Laboratory of Zoology, Occidental College, 1600 Campus Rd., Los Angeles, CA 90041 Phone: 323-259-1352 E-mail: mccormack@oxy.edu

ABSTRACT

Evolutionary relationships among birds in Neoaves, the clade comprising the vast majority of avian diversity, have vexed systematists due to the ancient, rapid radiation of numerous lineages. We applied a new phylogenomic approach to resolve relationships in Neoaves using target enrichment (sequence capture) and high-throughput sequencing of ultraconserved elements (UCEs) in avian genomes. We collected sequence data from UCE loci for 32 members of Neoaves and one outgroup (chicken) and analyzed data sets that differed in their amount of missing data. An alignment of 1,541 loci that allowed missing data was 87% complete and resulted in a highly resolved phylogeny with broad agreement between the Bayesian and maximum-likelihood (ML) trees. Although results from the 100% complete matrix of 416 UCE loci were similar, the Bayesian and ML trees differed to a greater extent in this analysis, suggesting that increasing from 416 to 1,541 loci led to increased stability and resolution of the tree. Novel results of our study include surprisingly close relationships between phenotypically divergent bird families, such as tropicbirds (Phaethontidae) and the sunbittern (Eurypygidae) as well as between bustards (Otididae) and turacos (Musophagidae). This phylogeny bolsters support for monophyletic waterbird and landbird clades and also strongly supports controversial results from previous studies, including the sister relationship between passerines and parrots and the non-monophyly of raptorial birds in the hawk and falcon families. Although significant challenges remain to fully resolving some of the deep relationships in Neoaves, especially among lineages outside the waterbirds and landbirds, this study suggests that increased data will yield an increasingly resolved avian phylogeny.

The diversification of modern birds occurred extremely rapidly, with all major orders and most families becoming distinct within a short window of 0.5 to 5 million years around the Cretaceous-Tertiary boundary [1-4]. As with other cases of ancient, rapid radiation, resolving deep evolutionary relationships in birds has posed a significant challenge. Some authors have hypothesized that the initial splits within Neoaves might be a hard polytomy that will remain irresolvable even with expanded data sets (reviewed in [5]). However, several recent studies have suggested that expanded genomic and taxonomic coverage will lead to an increasingly resolved avian tree of life [2,6,7].

Using DNA sequence data to reconstruct rapid radiations like the Neoaves phylogeny presents a practical challenge on several fronts. First, short speciation intervals provide little time for substitutions to accrue on internal branches, reducing the phylogenetic signal for rapid speciation events. Traditionally, the solution to this problem has been to collect additional sequence data, preferably from a rapidly evolving molecular marker such as mitochondrial DNA [8]. However, rapidly evolving markers introduce a new set of problems to the inference of ancient radiations: through time, substitutions across rapidly evolving markers overwrite older substitutions, resulting in signal saturation and homoplasy [9]. To address this challenge, some researchers have inferred ancient phylogeny using rare genomic changes, like retroposon insertions and indels, because rare changes are unlikely to occur in the same way multiple times, thereby minimizing homoplasy [10,11]. Though successful in some cases [12], retroposons are often insufficiently numerous to fully resolve relationships between taxa that rapidly radiated [13], and although often billed as being homoplasy-free, we now know that shared retroposon insertions can be due to independent events [14].

A second challenge to reconstructing ancient, rapid radiations is the randomness inherent to the process of gene sorting (i.e., coalescent stochasticity), which occurs even when gene histories are estimated with 100% accuracy [15]. The amount of conflict among gene-tree topologies due to coalescent stochasticity increases as speciation intervals get shorter [16]. Hemiplasy refers to gene-tree discord deep in phylogenies resulting from stochastic sorting processes that occurred long ago, but where the alleles are now fully sorted [17]. Accounting for hemiplasy requires increasing the number of loci interrogated and analyzing the resulting sequence data using species-tree methods that accommodate discordant gene histories [18-20].

Despite these challenges, our understanding of Neoaves phylogeny has steadily improved as genomic coverage and taxonomic coverage have increased [21]. Hackett et al. [6] – based on 169 species and 19 loci – provided a more resolved phylogeny of all birds than ever before. Combined with other studies during the previous decade, we now have a resolved backbone for the avian tree of life, including three well-supported clades: Neoaves, Palaeognathae (e.g., ostrich, emu, tinamous) and Galloanserae (e.g., ducks and chickens) [2,6,22-25]. Nonetheless, many relationships within Neoaves remain challenging to resolve despite the application of molecular tools such as whole mitochondrial genomes [26-28] and rare genomic changes [12-14,29]. Specifically, many of the basal nodes and the evolutionary affinities of enigmatic lineages (e.g., tropicbirds, hoatzin, sunbittern/kagu) within Neoaves continue to be poorly supported even when addressed with large data sets comprising a variety of molecular markers. This raises the question: Are there certain relationships deep in the Neoaves phylogeny that cannot be resolved regardless of the scope of the data collected?

Here, we apply a new method for collecting large amounts of DNA sequence data to address evolutionary relationships in Neoaves. This method, which involves simultaneous capture and high-throughput sequencing of hundreds of loci, addresses the main challenges of resolving ancient, rapid radiations – and is applicable throughout the tree of life. The markers we target are anchored by ultraconserved elements (UCEs), which are short stretches of highly conserved DNA. UCEs were originally discovered in mammals [30], but are also found in a wide range of other organisms [31-33]. UCEs allow for the convenient isolation and capture of independent loci among taxonomically distant species while providing phylogenetic signal in flanking regions [33,34]. Because variation in the flanks increases with distance from the core UCE, these markers display a balance between having a high enough substitution rate while minimizing saturation, providing information for estimating phylogenies at multiple evolutionary timescales [33,35]. UCEs are rarely found in duplicated genomic regions [36], making the determination of orthology more straightforward than in other markers (e.g., exons) or whole genomes, and UCEs are numerous among distantly related taxa, facilitating their use as discrete loci in species-tree analysis [33,35]. We employed sequence capture (i.e., bait-capture or target enrichment) to collect UCE sequence data from genomic DNA of 32 non-model bird species (Fig. 1) and used outgroup UCE data from the chicken genome to reconstruct evolutionary relationships in Neoaves.

## METHODS

We extracted DNA from tissue samples of 32 vouchered museum specimens (Table 1; Fig. 1), each from a different family within the traditional Neoaves group [37], using a phenol-chloroform protocol [38]. All samples for this project were loaned by, and used with permission of, the Louisiana State University Museum of Natural Science. We prepared sequencing libraries from purified DNA using Nextera library preparation kits (Epicentre Biotechnologies, Inc.), incorporating modifications to the protocol outlined in Faircloth et al. [33]. Briefly, following limited-cycle (16-19 cycles) PCR to amplify libraries for enrichment and concentration of amplified libraries to 147 ng/$\mu$L using a Speed-Vac, we individually enriched libraries for 2,386 UCE loci using 2,560 synthetic RNA capture probes (MyBaits, Mycroarray, Inc.). We designed capture probes targeting UCE loci that had high sequence identity between lizards and birds because previous work indicated that UCE loci from this set were useful for deep-level avian phylogenetics [33]. Following enrichment, we incorporated a custom set of indexed, Nextera adapters to each library [39] using enriched product as template in a limited-cycle PCR (16 cycles), and we sequenced equimolar pools of enriched, indexed libraries using 1 ½ lanes of single-end, 100 bp sequencing on an Illumina Genome Analyzer IIx (LSU Genomics Facility). The LSU Genomics Facility demultiplexed pooled reads following the standard Illumina pipeline, and we combined demultiplexed reads from each run for each taxon prior to adapter trimming, quality filtering, and contig assembly.

We filtered reads for adapter contamination, low-quality ends, and ambiguous bases using an automated pipeline (https://github.com/faircloth-lab/illumiprocessor) that incorporates Scythe (https://github.com/vsbuffalo/scythe) and Sickle (https://github.com/najoshi/sickle). We assembled reads for each taxon using Velvet v1.1.04 [40] and VelvetOptimiser v2.1.7 (S Gladman; http://bioinformatics.net.au/software.shtml), and we computed coverage across UCEs using tools from the AMOS package, as described in [33]. We used the PHYLUCE software package (https://github.com/faircloth-lab/phyluce; version m1.0-final) to align assembled contigs back to their associated UCE loci, remove duplicate matches, create a taxon-specific database of contig-to-UCE matches, and include UCE loci from the chicken (*Gallus gallus*) genome as outgroup sequences. We then generated two alignments across all taxa: one containing no missing data (i.e., all loci required to be present in all taxa) and one allowing up to 50% of the species to have data missing for a given locus. We built alignments using MUSCLE [41]. The steps specific to this analysis are available from https://gist.github.com/47e03463db0573c4252f.

For both alignments (missing data and no missing data), we prepared a concatenated alignment for MrBayes v3.1.2 [42] by estimating the most-likely finite-sites substitution model for individual UCE loci. Using a parallel implementation of MrAIC from the PHYLUCE package, we selected the best-fitting substitution model for all loci using AICc, and we grouped loci having the same substitution model into partitions. We assigned the parent substitution model to each partition, for a total of 20 partitions, and we analyzed these alignments using two independent MrBayes runs (4 chains) of 10M iterations each (thinning=100). We sampled 50,000 trees from the posterior distribution (burn-in=50%) after convergence by ensuring the average standard deviation of split frequencies was $< 0.00001$ and the potential scale reduction factor for estimated parameters was approximately 1.0. We confirmed convergence with Effective Sample Size values $>200$ in TRACER [43] and by assessing the variance in tree topology with AWTY [44]. We also prepared a concatenated alignment in PHYLIP format with a single partition containing all sequence data, and we analyzed this alignment using the fast-approximation, maximum likelihood (ML) algorithm in RaXML (raxmlHPC-MPI-SSE3; v. 7.3.0) with 1,000 bootstrap replicates [45,46].

For the data set with no missing data, we also estimated a species tree on 250 nodes of a Hadoop cluster (Amazon Elastic Map Reduce) using a map-reduce implementation (https://github.com/ngcrawford/CloudForest) of a workflow combining MrAIC to estimate and select the most-appropriate finite-sites substitution model. We used PhyML 3.0 [47] to estimate gene trees, and PHYBASE to estimate species trees from gene trees using the STAR (Species

Trees from Average Ranks of Coalescences) method [48]. We performed 1,000 multi-locus, non-parametric bootstrap replicates for the STAR tree by resampling nucleotides within loci as well as resampling loci within the data set [49]. We only performed the species tree analysis on the alignment with no missing data due to concerns about how missing loci might affect a coalescent analysis.

To assess phylogenetically informative indels, we scanned alignments by eye in Geneious 5.4 (Biomatters Ltd, Aukland, New Zealand), recording indels that were 2 bp or more in length and shared between two or more ingroup taxa. We then mapped informative indels onto the resolved 416-locus Bayesian phylogeny.

## RESULTS

We provide summary statistics for sequencing and alignment in Table 1. We obtained an average of 2.6 million reads per sample (range = 1.1 – 4.9 million). These reads assembled into an average of 1,830 contigs per sample (range = 742 – 2,418). An average (per sample) of 1,412 of these contigs matched the UCE loci from which we designed target capture probes (range = 694 – 1,681). The average length of UCE-matching contigs was 429 base pairs (bp) (range = 244 – 598), and the average coverage of UCE-matching contigs was 71 times (range = 44 – 138). The percentage of original sequencing reads that were "on target" (i.e., helped build UCE-matching contigs) averaged 24% across samples (range = 15% - 35%).

When we selected loci allowing 50% of species for a given locus to have missing data, the final data set contained 1,541 UCE loci and produced a concatenated alignment that was 87% complete across 32 Neoaves species and the chicken outgroup. The average length of these 1,541 loci was 350 bp (min=90, max=621), and the total concatenated alignment length was 539,526 characters (including indels) with 24,703 informative sites.

Generally, the Bayesian and ML phylogenies for the 1,541 locus alignment were similar in their topology and amount of resolution (Fig. 2a; see Fig. S1 for fully resolved trees). Of the 31 nodes, 27 (87%) were highly supported in the Bayesian tree (>0.95 PP), whereas a subset of 20 of those nodes (65%) were also highly supported in the ML tree (>75% bootstrap score). An additional 7 nodes (23%) appeared in both the Bayesian and ML trees, but support in the ML tree was low (bisected nodes in Fig. 2a). Four nodes (16%) had either low support in both trees (and thus are collapsed in Fig. 2a) or had high support in the Bayesian tree, but did not appear in the ML tree (white nodes in Fig. 2a). A phylogram for the 1,541 locus Bayesian tree (Fig. S2) showed long terminal branches and short internodes near the base of the tree, consistent with previous studies suggesting an ancient, rapid radiation of Neoaves.

For the data set requiring no missing data, we recovered 416 UCE loci across 29 Neoaves species and the chicken outgroup. Enrichments for three species performed relatively poorly (Table 1; *Micrastur*, *Trogon*, and *Vidua*), and we excluded these samples to boost the number of loci recovered. The average length of these 416 loci was 397 bp, and the total concatenated alignment length was 165,163 characters (including indels) with 7,600 informative sites. Bayesian and ML trees differed more in their topology and resolution than was observed for the 1,541 locus trees above (Fig. 2b; see Fig. S3 for fully resolved trees). Of the 28 nodes, 24 (86%) were highly supported in the Bayesian tree (>0.95 PP), whereas only a subset of 14 (50%) was highly supported in the ML tree (>75% bootstrap score). We recovered an additional three nodes (11%) in both the Bayesian and ML trees, but support for these nodes in the ML tree was low (bisected nodes in Fig. 2b). Twelve nodes (43%) disagreed between the Bayesian and ML trees, a frequency much higher than the 16% disagreement we observed from the 1,541 locus analysis.

The STAR species tree from the 416 locus data set (Fig. 3; Fig. S3c) was much less resolved and had lower support values than either the Bayesian or ML tree estimated for these data. There has been little study on what constitutes high bootstrap support for a species tree analysis, but only 11 nodes (39%) had over 50% support. Despite the differences in resolution between the Bayesian, ML, and STAR species tree for the 416 locus analysis, when we collapsed weakly supported nodes (PP < 0.90, ML bootstrap < 70%, species-tree bootstrap < 40%), there very few strongly supported contradictions among the three trees.

We identified 44 indels greater than two bp in length that were shared among two or more ingroup taxa (Table S1). Only 13 of these indels validated clades found in the phylogenetic trees generated from nucleotide data. The four clades supported by the 13 indels represented four of the six longest internal branches of the phylogeny (Fig. 4).

DISCUSSION

Containing 1,541 loci and 32 species, our study is among the largest comparative avian phylogenomics data sets assembled for the purpose of elucidating avian evolutionary relationships. By strengthening support for controversial relationships and resolving several new parts of the avian tree (discussed below), our results suggest that increasing sequence data will lead to an increasingly resolved bird tree of life, with some caveats. Our sampling strategy sought to balance the number of taxa included with the number of loci interrogated. We sampled the genome much more broadly than the 19 loci of Hackett et al. [6], but with reduced taxonomic sampling (32 species compared to 169 species). Additionally, compared to Hackett et al. [6], our loci were shorter (350 bp vs. 1,400 bp), meaning that although our 1,541 locus data set contained roughly 80 times the number of loci, our total alignment length was only about 17 times larger.

Another recent avian phylogenomic study [50] included 1,995 loci, producing a concatenated alignment roughly 1.5 times larger than ours, but this study included only 9 Neoaves species, 5 of which were passerines, which limited the potential of that study for phylogenetic inference.

*Increasing data increases resolution of the avian tree of life*

One striking result of our study is that Bayesian and ML trees based on 1,541 loci were in much stronger agreement with one another than Bayesian and ML trees estimated from 416 loci (Fig. 2). The stronger agreement was driven primarily by increased resolution and support of the 1,541 locus ML tree (i.e., it became more similar to the Bayesian tree). In contrast, although the 416-locus Bayesian tree was highly resolved, its ML counterpart was much less so and conflicted in topology with the Bayesian tree to a greater degree.

Combined with results of other studies, this suggests that increasing loci leads to increasing support and stability of the avian tree. In discussing our results below, we rely primarily on relationships found in the 1,541 locus tree due to the stronger congruence among analytical methods, as well as recent research suggesting that analyses of incomplete data matrices may be beneficial for studies with highly incomplete taxonomic sampling [51]. Most simulation studies assessing the effect of missing data found that a common negative effect of missing data was erosion of support values rather than an artificial increase in support [52]. We did not observe lower support values in the tree with more missing data, and, in fact, we observed the opposite, suggesting minimal negative effects of missing data. This is perhaps unsurprising given that the threshold amount of missing data producing negative effects in simulation studies was often much higher than our level of missing data (many studies assessing 50-90% missing data, whereas we had 13%). Where relevant, we compare the 416 locus tree and species tree to the 1,541 locus tree, and we discuss a few results from the 416 locus tree that are particularly well supported or interesting.

*Low support for the species tree and differences between Bayesian and ML trees*

The low support for many nodes in the species tree (Fig. 3) is understandable given the length of individual UCE loci. We estimated the species tree using methods that take gene trees as input, rather than those that jointly estimating both gene trees and species trees [53], which is too computationally intensive for large data sets. Therefore, the resolution of the species tree is entirely dependent on the quality and resolution of the individual gene trees. Because we assembled relatively short UCE loci (397 bp for the 416 locus data set) from enriched reads, each locus, considered individually, is not likely to contain much signal informing basal relationships. Concatenation effectively masks this reduction in signal by joining all loci, maximizing the information content on short internal branches, and helping to resolve relationships when speciation intervals are short. Of course, this benefit of concatenation comes with the cost of ignoring the independent histories of genes and potentially inflating support values for nodes affected by substantial coalescent stochasticity [54,55], especially when using Bayesian methods.

While the low information content of shorter UCE loci clearly posed a problem for inferring the species tree, this is a methodological limitation of this study rather than a general limitation of the UCE enrichment approach. For this study, we sequenced single-end, 100 bp reads on an Illumina GAIIx.  However, it is now possible to obtain paired-end reads as long as 250 bp from the Illumina platform, which will facilitate assembly of longer loci from fewer reads than we obtained during this study. Tighter control on the average size of DNA fragments used for enrichment (i.e., using fragments of the maximum size allowed by the sequencing platform) and increased sequencing depth can also increase the size of recovered loci to 600-700 bp (B. Faircloth, unpublished data). Using UCE loci that averaged ~750 bp, we did not observe poorly resolved species trees in a study of rapid radiation of mammals [35]. Thus, increasing the length of loci recovered is clearly an important step towards addressing the dual problems of low information content and coalescent stochasticity in resolving the avian tree of life, although it remains to be seen how denser taxon sampling will interact with these problems and affect future analyses. In any event, given our results and those of prior studies, the more exigent problem in this case appears to be low information content.

Although there were very few contradictory relationships in highly supported parts of the trees, there was an obvious difference in resolution between the Bayesian and ML trees for the 416 locus alignment, and to a lesser degree, for the 1,541 locus alignment. One possible explanation for the lower resolution of the ML trees is that bootstrapping may not be the best way to assess confidence with UCE data, given the expected skewed distribution of phylogenetic information across sites (i.e., more toward the flanks) [33]. Also, it is common to observe higher support values for trees estimated by Bayesian methods, and in some cases PPs can be deceptively high [56,57]. There is also current debate concerning whether Bayesian methods might suffer from a "star tree paradox", where a simultaneous divergence of three or more lineages nonetheless appears resolved in bifurcating fashion with high PP [58,59]. Bayesian methods also might be more prone to long-branch attraction [60]. Research on these concerns is ongoing and salient to our results, in which the Bayesian trees tended to group several basally diverging lineages with long branches together into clades with high PP that were not supported by the ML trees. On the other hand, ML bootstraps can underestimate support compared to Bayesian methods [61,62] – an effect suggested by our observation that many weakly supported nodes in the 416 locus ML tree, for which Bayesian analysis showed high PP, became well supported in the ML tree when we increased the size of the data matrix to 1,541 loci.

*Defining a backbone for the Neoaves phylogeny*

We found strong congruence across data sets and analytical methods for previously hypothesized, but still tenuously supported, waterbird (Aequornithes; [63]) and landbird clades [2,6] that diverge deep in the Neoaves phylogeny (Fig. 2). We address relationships within landbirds and waterbirds below, but their position as sister clades in three of four trees contrasts with previous studies that placed a number of additional taxa close to the waterbirds [2,6,23]. Both Bayesian trees supported a third clade – including families as diverse as hummingbirds, flamingos, cuckoos, trumpeters, bustards, and turacos – bearing some resemblance to the Metaves clade recovered in earlier molecular studies [2,6,23], but differing by including bustards, trumpeters, and turacos, which have not typically been considered part of Metaves. However, this clade did not appear in either ML tree or the species tree, suggesting that the grouping of these taxa could be an artifact resulting from long-branch attraction, as discussed above. Although we uncovered novel, well-supported sister relationships between some of these species toward the tips of the tree (see below), their deeper evolutionary affinities will need to be explored with increased taxonomic sampling to break up long branches and provide further information on state changes deep in the tree. Our study thus suggests that resolving the avian tree outside of waterbirds and landbirds is the final frontier in deep-level bird systematics.

*The surprising relationship between tropicbirds and the sunbittern*

This study adds to the overwhelming evidence for a sister relationship between the phenotypically divergent flamingo and grebe families [2,5,6,64-66]. Our results also suggest another surprisingly close affinity between morphologically disparate groups – tropicbirds and the sunbittern. Three of four analyses lent strong support to this relationship, for which ML support increased sharply (43% to 96%) when genomic sampling increased from 416 to 1,541 loci (Fig. 2; Fig. S1 & S2). A close relationship between the sunbittern and tropicbirds is surprising because of dissimilarities in appearance, habitat, and geography. Tropicbirds are pelagic seabirds with mostly white plumage, elongated central tail feathers, and short legs that make walking difficult. Meanwhile, the sunbittern is a cryptic resident of lowland and foothill Neotropical forests that spends much of its time foraging on the ground in and near freshwater streams and rivers. The kagu, a highly terrestrial bird restricted to the island of New Caledonia (not sampled in our study), is the sister species of the sunbittern [6,22,23] and may superficially bear some similarity to tropicbirds. These results should spark further research into shared morphological characteristics of tropicbirds, the sunbittern, and the kagu.

*A sister relationship between bustards and turacos?*

Another surprising sister relationship uncovered in our study is that between turacos and bustards (Fig. 2a). Turacos are largely fruit-eating arboreal birds of sub-Saharan Africa, whereas bustards are large, omnivorous, terrestrial birds widely distributed in the Old World. Despite some overlap in their biogeography, the two families have little in common and have, to our knowledge, never been hypothesized to be closely related based on phenotypic characteristics. Previous molecular studies have placed members of these two families near one another evolutionarily [2,6], but never as sister taxa. Our study did not include a member of the cuckoo family, which has often been considered a close relative of the turacos and thus might be its true sister taxon. An additional note of caution is that a turaco-bustard relationship was not supported outside the 1,541 locus tree, but neither was it contradicted. Thus, although confirming results are needed, our study provides some support for the idea that turacos and bustards are much more closely related than previously thought, if not actually sister families.

*Further clarity for waterbird relationships*

We found consistent support across all analyses for relationships among the six sampled families within the waterbirds (Figs. 2 and 3). Prior to the availability of molecular data, the relationships within this clade were difficult to resolve due to the extreme morphological diversity of its members and the scarcity of apomorphic morphological characters [63]. The topology we recovered within this portion of the tree is identical to that of Hackett et al. [6]. For example, in both studies loons are the outgroup to all other waterbirds, and the morphologically divergent penguins are sister to tube-nosed seabirds in the family Procellariidae.

*Hoatzin: still a riddle wrapped in a mystery…*

Hoatzin (*Opisthicomus hoazin*), the only extant member of Opisthocomidae, is arguably the most enigmatic living bird species due to its unique morphology, folivorous diet, and confusion relative to its evolutionary affinities across numerous molecular phylogenies. One phylogenetic study found no support for a sister relationship between hoatzin and the Galloanserae, nor with turacos, cuckoos, falcons, trogons, or mousebirds in Neoaves; the study found some, albeit weak, support for a sister relationship between hoatzin and doves [67]. The 416 locus Bayesian tree placed the hoatzin sister to a shorebird (Fig. 2b) with high support, but we did not observe this relationship in either the ML tree or the species tree. Furthermore, support for any definitive placement of the hoatzin eroded in the 1,541 locus tree (Fig. 2a). A close relationship of hoatzin to shorebirds would be extremely surprising and in stark contrast to any prior hypotheses [68]. Our results raise the question of whether or not more data will eventually lead to a definitive conclusion on the phylogenetic position of the hoatzin. Given the phylogenetic distinctiveness of the hoatzin, better taxonomic sampling may be as beneficial as further genomic sampling in the search for shared, derived characters deep in the tree. Thus, we present a link between the hoatzin and shorebirds, a large family whose members are found in diverse terrestrial and aquatic habitats, as an intriguing phylogenetic hypothesis.

*An early divergence for pigeons and doves?*

Another place where our 416 locus trees showed support for a relationship not found in the 1,541 locus trees was in the placement of the pigeon and dove family (Columbidae). Most prior studies either placed pigeons and doves in an unresolved position [6] or sister to sandgrouse (Pteroclididae) within Metaves [2]. However, amino acid sequences of feather beta-keratins have suggested a basal position of Columbidae within Neoaves [69]. We found complete support in the 416-locus Bayesian tree for a sister relationship between Columbidae and the rest of Neoaves (Fig. 2b). We also recovered this relationship in the 416-locus ML tree and species tree, although with weak support (Fig. S2). However, the 1,541 locus trees disagreed by placing pigeons and doves in a more conventional position sister to sandgrouse and instead placing trumpeters sister to the rest of Neoaves (Fig. 2a).

*Support for controversial relationships within the landbirds*

One of the biggest challenges to conventional thought on bird phylogeny contained in Hackett et al. [6] was in the relationships among landbirds. Their finding that parrots were the sister family to passerines is still viewed as controversial (bootstrap support for parrots + passerines from Hackett et al. [6] was 77%), despite corroborating evidence from rare genomic changes encoded in retroposons [12] and expanded data sets [7]. Our results across all analyses strongly support the sister relationship between passerines (in this study represented by a suboscine *Pitta* and an oscine *Vidua*) and parrots (perfect support in all Bayesian and ML trees; 85% support in the species tree).

Our results also support another controversial finding from Hackett et al. [6]: the absence of a sister relationship between raptorial birds in the hawk (Accipitridae) and falcon (Falconidae) families. Both ML and Bayesian trees from the 1,541 locus analysis provided perfect support for falcons sister to the parrot + passerine clade, whereas the representative of the hawk family was sister to the vultures with high support, improving upon the weak support for hawks + vultures from Hackett et al. [6].

Finally, the larger 1,541 analysis helped resolve deeper relationships within the landbirds among four main clades: (i) passerines + parrots + falcons, (ii) hawks + vultures, (iii) the group sometimes called the "near passerines" (e.g., barbet, woodpecker, woodhoopoe, motmot, and trogon, also known as the CPBT clade in [7] because it includes the families Coraciiformes, Piciformes, Bucerotiformes, and Trogoniformes), and (iv) owls (Fig. 2a). The Bayesian tree placed owls sister to the "near passerines" and then hawks + vultures sister to owls + "near passerines", a topology that also appeared in the ML tree with weak support.

Meanwhile, the evolutionary affinities of mousebirds, whose position in prior studies has been uncertain [6,7], remain equivocal. The 416 locus trees positioned mousebirds sister to the "near passerines", but the 1,541 locus trees placed mousebirds sister to passerines. Wang et al. [7] also found mousebirds moving between these two clades depending on the analysis. Other relationships within the "near passerines" were consistent with previous results [2,6] except that the positions of trogons and motmots switched between the 416 and 1,541 locus trees.

*A scarcity of indels on short internal branches*

Our finding that informative indels were generally scarce (found only on four of the longest internal branches in the phylogeny; Fig. 4) corroborates previous work on rare genomic changes in retroposons, which also found little evidence for shared events deep in the bird phylogeny [12,13]. The low prevalence of informative indels may be exacerbated by the lack of major structural changes in and around UCE loci, although this has not been well studied. Previous work on nuclear introns has identified a handful of indels supporting major subdivisions deep in avian phylogeny [23,70,71]. However, lessons from coalescence theory caution that, when drawing phylogenetic inferences from rare genomic changes, numerous loci supporting particular subdivisions are required to account for the expected high variance in gene histories [35]. The study of bird phylogeny awaits a genome-scale analysis of many hundreds of rare genomic events including indels, retroposons, and microRNAs.

*Conclusions*

Our results, combined with other recent studies [2,6], demonstrate that increasing sequence data leads to improved resolution of the bird tree of life. Major challenges clearly remain in corroborating results across analytical methods and data types. One of these challenges is a species tree for birds. While we have focused here on the seemingly more pressing problem of obtaining phylogenetic signal and high support values from concatenated data sets, we acknowledge that a proper accounting of the ultra-rapid radiation of avian lineages will require methods that reconcile discordant gene trees, which could lead to different results. Nevertheless, the incremental progress of resolving the bird tree of life is a major turnaround from more pessimistic attitudes that predated the decreased sequencing costs of the last decade and the advent of high-throughput sequencing technologies [72].

The framework we outline here, sequence capture using UCEs, is a powerful approach that can scale to hundreds of taxa, thousands of loci, and include longer flanking sequences with different library preparation and sequencing regimes. Because UCEs occur in many organisms, the method is broadly applicable across the tree of life [32,33]. Data from sequence capture approaches can also be mixed, in hybrid fashion, with UCEs excised from whole genome assemblies [33,34,73] or other types of molecular markers, providing a powerful method for collecting and analyzing phylogenomic data from non-model species to elucidate their evolutionary histories.

## LITERATURE CITED

1. Feduccia A (1999) The origin and evolution of birds. New Haven, Connecticut: Yale University Press.

2. Ericson PGP, Anderson CL, Britton T, Elzanowski A, Johansson US, et al. (2006) Diversification of Neoaves: integration of molecular sequence data and fossils. Biol Lett 2: 543-547.

3. Brown JW, Rest J, García-Moreno J, Sorenson M, Mindell D (2008) Strong mitochondrial DNA support for a Cretaceous origin of modern avian lineages. BMC Biol 6: 6.

4. Chojnowski JL, Kimball RT, Braun EL (2008) Introns outperform exons in analyses of basal avian phylogeny using clathrin heavy chain genes. Gene 410: 89-96.

5. Cracraft J, Barker FK, Braun M, Harshman J, Dyke GJ, et al. (2004) Phylogenetic relationships among modern birds (Neornithes): toward an avian tree of life. In: Cracraft J, Donoghue M, editors. Assembling the tree of life. New York, NY: Oxford University Press. pp. 468-489.
6. Hackett SJ, Kimball RT, Reddy S, Bowie RCK, Braun EL, et al. (2008) A phylogenomic study of birds reveals their evolutionary history. Science 320: 1763-1768.
7. Wang N, Braun EL, Kimball RT (2012) Testing hypotheses about the sister group of the Passeriformes using an independent 30-locus data set. Mol Biol Evol 29: 737-750.
8. Zink RM, Barrowclough GF (2008) Mitochondrial DNA under siege in avian phylogeography. Mol Ecol 17: 2107-2121.
9. Whitfield JB, Lockhart PJ (2007) Deciphering ancient rapid radiations. Trends Ecol Evol 22: 258-265.
10. Rokas A, Holland PWH (2000) Rare genomic changes as a tool for phylogenetics. Trends Ecol Evol 15: 454-459.
11. Shedlock AM, Takahashi K, Okada N (2004) SINEs of speciation: tracking lineages with retroposons. Trends Ecol Evol 19: 545-553.
12. Suh A, Paus M, Kiefmann M, Churakov G, Franke FA, et al. (2011) Mesozoic retroposons reveal parrots as the closest living relatives of passerine birds. Nature Comm 2: 443.
13. Matzke A, Churakov G, Berkes P, Arms EM, Kelsey D, et al. (2012) Retroposon insertion patterns of neoavian birds: strong evidence for an extensive incomplete lineage sorting era. Mol Biol Evol 29: 1497-1501.
14. Han KL, Braun EL, Kimball RT, Reddy S, Bowie RCK, et al. (2011) Are transposable element insertions homoplasy free?: an examination using the avian tree of life. Syst Biol 60: 375-386.
15. Maddison WP (1997) Gene trees in species trees. Syst Biol 46: 523-536.
16. Degnan JH, Salter LA (2005) Gene tree distributions under the coalescent process. Evolution 59: 24-37.
17. Avise JC, Robinson TJ (2008) Hemiplasy: a new term in the lexicon of phylogenetics. Syst Biol 57: 503-507.
18. Knowles L (2009) Estimating species trees: methods of phylogenetic analysis when there is incongruence across genes. Syst Biol 58: 463-467.

19. Edwards SV (2008) Is a new and general theory of molecular systematics emerging? Evolution 63: 1-19.
20. Liu L, Yu L, Kubatko LS, Pearl DK, Edwards SV (2009) Coalescent methods for estimating phylogenetic trees. Mol Phylogenet Evol 53: 320-328.
21. Edwards SV, Jennings WB, Shedlock AM (2005) Phylogenetics of modern birds in the era of genomics. Proc R Soc B 272: 979.
22. Livezey BC, Zusi RL (2007) Higher-order phylogeny of modern birds (Theropoda, Aves: Neornithes) based on comparative anatomy. II. Analysis and discussion. Zool J Linn Soc 149: 1-95.
23. Fain MG, Houde P (2004) Parallel radiations in the primary clades of birds. Evolution 58: 2558-2573.
24. Groth JG, Barrowclough GF (1999) Basal divergences in birds and the phylogenetic utility of the nuclear RAG-1 gene. Mol Phylogenet Evol 12: 115-123.
25. Haddrath O, Baker AJ (2012) Multiple nuclear genes and retroposons support vicariance and dispersal of the palaeognaths, and an Early Cretaceous origin of modern birds. Proc Roy Soc B In press.
26. Gibb GC, Penny D (2010) Two aspects along the continuum of pigeon evolution: A South-Pacific radiation and the relationship of pigeons within Neoaves. Mol Phylogenet Evol 56: 698-706.
27. Pratt RC, Gibb GC, Morgan-Richards M, Phillips MJ, Hendy MD, et al. (2009) Toward resolving deep Neoaves phylogeny: data, signal enhancement, and priors. Mol Biol Evol 26: 313-326.
28. Pacheco MA, Battistuzzi FU, Lentino M, Aguilar RF, Kumar S, et al. (2011) Evolution of modern birds revealed by mitogenomics: timing the radiation and origin of major orders. Mol Biol Evol 28: 1927-1942.
29. Braun E, Kimball R, Han KL, Iuhasz-Velez N, Bonilla A, et al. (2011) Homoplastic microinversions and the avian tree of life. BMC Evol Biol 11: 141.
30. Bejerano G, Pheasant M, Makunin I, Stephen S, Kent WJ, et al. (2004) Ultraconserved elements in the human genome. Science 304: 1321-1325.

31. Janes DE, Chapus C, Gondo Y, Clayton DF, Sinha S, et al. (2011) Reptiles and mammals have differentially retained long conserved noncoding sequences from the Amniote ancestor. Genome Biol Evol 3: 102-113.

32. Siepel A, Bejerano G, Pedersen JS, Hinrichs AS, Hou M, et al. (2005) Evolutionarily conserved elements in vertebrate, insect, worm, and yeast genomes. Genome Res 15: 1034-1050.

33. Faircloth BC, McCormack JE, Crawford NG, Harvey MG, Brumfield RT, et al. (2012) Ultraconserved elements anchor thousands of genetic markers for target enrichment spanning multiple evolutionary timescales. Syst Biol 61: 717-726.

34. Crawford NG, Faircloth BC, McCormack JE, Brumfield RT, Winker K, et al. (2012) More than 1000 ultraconserved elements provide evidence that turtles are the sister group of archosaurs. Biol Lett 8: 783-786.

35. McCormack JE, Faircloth BC, Crawford NG, Gowaty PA, Brumfield RT, et al. (2012) Ultraconserved elements are novel phylogenomic markers that resolve placental mammal phylogeny when combined with species-tree analysis. Genome Res 22: 746-754.

36. Derti A, Roth FP, Church GM, Wu C-t (2006) Mammalian ultraconserved elements are strongly depleted among segmental duplications and copy number variants. Nat Genet 38: 1216-1220.

37. Sibley CG, Monroe BL (1990) Distribution and taxonomy of birds of the world. New Haven: Yale University Press.

38. Sambrook J, Russell DW (2001) Molecular cloning: a laboratory manual. Cold Spring Harbor: CSHL Press.

39. Faircloth BC, Glenn TC (2012) Not all sequence tags are created equal: designing and validating sequence identification tags robust to indels. PLoS One 7: e42543.

40. Zerbino DR, Birney E (2008) Velvet: Algorithms for de novo short read assembly using de Bruijn graphs. Genome Res 18: 821-829.

41. Edgar RC (2004) MUSCLE: a multiple sequence alignment method with reduced time and space complexity. BMC Bioinformatics 5: 113-119.

42. Ronquist F, Huelsenbeck JP (2003) MrBayes 3: Bayesian phylogenetic inference under mixed models. Bioinformatics 19: 1572-1574.

43. Rambaut A, Drummond AJ (2007) Tracer - MCMC Trace Analysis Tool, v1.4. Available from: <http://beastbioedacuk/Tracer/>.

44. Nylander JAA, Wilgenbusch JC, Warren DL, Swofford DL (2008) AWTY (are we there yet?): a system for graphical exploration of MCMC convergence in Bayesian phylogenetics. Bioinformatics 24: 581-583.

45. Stamatakis A (2006) RAxML-VI-HPC: maximum likelihood-based phylogenetic analyses with thousands of taxa and mixed models. Bioinformatics 22: 2688-2690.

46. Stamatakis A, Hoover P, Rougemont J (2008) A rapid bootstrap algorithm for the RAxML Web servers. Syst Biol 57: 758-771.

47. Guindon S, Dufayard JF, Lefort V, Anisimova M, Hordijk W, et al. (2010) New algorithms and methods to estimate maximum-likelihood phylogenies: assessing the performance of PhyML 3.0. Syst Biol 59: 307-321.

48. Liu L, Yu L, Pearl DK, Edwards SV (2009) Estimating species phylogenies using coalescence times among sequences. Syst Biol 58: 468-477.

49. Seo TK (2008) Calculating bootstrap probabilities of phylogeny using multilocus sequence data. Mol Biol Evol 25: 960-971.

50. Nabholz B, Kunstner A, Wang R, Jarvis ED, Ellegren H (2011) Dynamic evolution of base composition: causes and consequences in avian phylogenomics. Mol Biol Evol 28: 2197-2210.

51. Wiens JJ, Tiu J (2012) Highly incomplete taxa can rescue phylogenetic analyses from the negative impacts of limited taxon sampling. PLoS One 7: e42925.

52. Wiens JJ, Morrill MC (2011) Missing data in phylogenetic analysis: reconciling results from simulations and empirical data. Syst Biol 60: 719-731.

53. Edwards SV, Liu L, Pearl DK (2007) High-resolution species trees without concatenation. Proc Natl Acad Sci USA 104: 5936-5841.

54. Kubatko L, Degnan J (2007) Inconsistency of phylogenetic estimates from concatenated data under coalescence. Syst Biol 56: 17-24.

55. Mossel E, Vigoda E (2005) Phylogenetic MCMC algorithms are misleading on mixtures of trees. Science 309: 2207-2209.

56. Suzuki Y, Glazko GV, Nei M (2002) Overcredibility of molecular phylogenies obtained by Bayesian phylogenetics. Proc Natl Acad Sci USA 25: 16138-16143.

57. Douady CJ, Delsuc F, Boucher Y, Doolittle WF, Douzery EJP (2003) Comparison of Bayesian and maximum likelihood bootstrap measures of phylogenetic reliability. Mol Biol Evol 20: 248-254.
58. Kolaczkowski B, Thornton JW (2006) Is there a star tree paradox? Mol Biol Evol 23: 1819-1823.
59. Lewis PO, Holder MT, Holsinger KE (2005) Polytomies and Bayesian phylogenetic inference. Syst Biol 54: 241-253.
60. Kolaczkowski B, Thornton JW (2009) Long-branch attraction bias and inconsistency in Bayesian phylogenetics. PLoS One 4: e7891.
61. Erixon P, Svennblad B, Britton T, Oxelman B (2003) Reliability of Bayesian posterior probabilities and bootstrap frequencies in phylogenetics. Syst Biol 52: 665-673.
62. Huelsenbeck JP, Larget B, Miller RE, Ronquist F (2002) Potential applications and pitfalls of Bayesian inference of phylogeny. Syst Biol 51: 673-688.
63. Mayr G (2011) Metaves, Mirandornithes, Strisores and other novelties: a critical review of the higher-level phylogeny of neornithine birds. J Zool Syst Evol Res 49: 58-76.
64. Chubb AL (2004) New nuclear evidence for the oldest divergence among neognath birds: the phylogenetic utility of ZENK (i). Mol Phylogenet Evol 30: 140-151.
65. van Tuinen M, Butvill DB, Kirsch JAW, Hedges SB (2001) Convergence and divergence in the evolution of aquatic birds. Proc R Soc Lond B 268: 1345-1350.
66. Morgan-Richards M, Trewick S, Bartosch-Härlid A, Kardailsky O, Phillips M, et al. (2008) Bird evolution: testing the Metaves clade with six new mitochondrial genomes. BMC Evol Biol 8: 20.
67. Sorenson MD, Oneal E, Garcia-Moreno J, Mindell DP (2003) More taxa, more characters: the hoatzin problem is still unresolved. Mol Biol Evol 20: 1484-1498.
68. Thomas B (1996) Family Opisthocomidae (hoatzins). In: del Hoyo J, Jordi A, Sargatal C, editors. Handbook of the birds of the world, volume 3, hoatzins to auks. Barcelona: Lynx Ediciones. pp. 24-32.
69. Glenn TC, French JO, Heincelman TJ, Jones KL, Sawyer RH (2008) Evolutionary relationships among copies of feather beta (β) keratin genes from several avian orders. Integr Comp Biol 48: 463-475.

70. Pasko L, Ericson PGP, Elzanowski A (2011) Phylogenetic utility and evolution of indels: A study in neognathous birds. Mol Phylogenet Evol 61: 760-771.

71. Prychitko TM, Moore WS (2003) Alignment and phylogenetic analysis of β-fibrinogen intron 7 sequences among avian orders reveal conserved regions within the intron. Mol Biol Evol 20: 762-771.

72. Poe S, Chubb AL (2004) Birds in a bush: five genes indicate explosive evolution of avian orders. Evolution 58: 404-415.

73. Haussler D, O'Brien SJ, Ryder OA, Barker FK, Clamp M, et al. (2009) Genome 10K: a proposal to obtain whole-genome sequence for 10,000 vertebrate species. J Hered 100: 659-674.

**Acknowledgments** We thank Scott Herke and the LSU Genomics Facility for assistance with sequencing. Donna Dittmann (LSUMZ) assisted with tissue loans. Illustrations for Figure 1 are artistic interpretations based on photos used with permission or under Creative Commons license. Photo credits for Fig. 1: (1) Prin Pattawaro; (2) Alan Manson; (3) Farelli; (4) Unknown; (5) Enoch Joseph Wetsy; (6) Vijay Cavale; (7) Nancy Wyman; (8) Lip Kee; (9) Steve Turner; (10) Keith Murdock; (11) Unknown; (12) Eduardo Lopez; (13) Jan Sevcik; (14) Srihari Kulkarni; (15) Arthur Grosset; (16) Tom Tarrant; (17) Fir0002/Flagstaffotos; (18) Mark Hannaford; (19) Unknown; (20) Jose Garcia; (21) "The Lilac Breasted Roller"; (22) Paul Baker; (23) Unknown; (24) Linda De Volder; (25) Utz Klingenböck; (26) Pixxl (Lisa M); (27) Bobby K; (28) Tarique Sani; (29) Lee Harding; (30) Doug Backlund; (31) Doug Pratt; (32) Jeff Whitlock. M. Alfaro and two anonymous reviewers provided comments on the manuscript.

**Availability** Assembled contigs, alignments, and gene trees for both data sets are available from Dryad (doi: 10.5061/dryad.sd080). All source code used for UCE data processing is available from https://github.com/faircloth-lab/phyluce under BSD and Creative Commons licenses. Version controlled, reference probe sets and outgroup data are available from https://github.com/faircloth-lab/uce-probe-sets. UCE contigs used in analyses are available from Genbank (accessions: JQ328245 - JQ335930, KC358654 - KC403881). Protocols for UCE enrichment, probe design, and additional information regarding techniques are available from http://ultraconserved.org.

## Table 1. Summary of descriptive statistics for samples, Illumina sequencing, and UCE loci.

| | | | | | All contigs | | | | Contigs Aligned to UCE loci | | | | | | |
|---|---|---|---|---|---|---|---|---|---|---|---|---|---|---|---|
| Family | Scientific name | Common Name | Museum tissue no. | Number of trimmed reads | Count | Avg. size | Avg. coverage | Reads in contigs | Count | Avg. size | Avg. coverage | Reads in contigs | Contigs match >1 locus[1] | Contigs "on-target"[2] | Reads "on-target"[3] |
| Pittidae (1) | *Pitta guajana* | Banded Pitta | LSUMZ B36368 | 2,723,264 | 2369 | 386 | 63.1 | 914,414 | 1572 | 457.4 | 71.3 | 719,095 | 32 | 0.66 | 0.26 |
| Viduidae (2) | *Vidua macroura* | Pin-tailed Whydah | LSUMZ B16749 | 1,098,154 | 1203 | 240 | 41.3 | 288,210 | 959 | 244.2 | 43.5 | 234,214 | 2 | 0.80 | 0.21 |
| Psittacidae (3) | *Psittacula alexandri* | Red-breasted Parakeet | LSUMZ B36554 | 2,745,979 | 2312 | 421 | 55 | 974,441 | 1487 | 508.1 | 62.7 | 752,493 | 42 | 0.64 | 0.27 |
| Falconidae (4) | *Micrastur* | Collared Forest Falcon | LSUMZ B11298 | 1,405,847 | 742 | 309 | 49.9 | 229,417 | 694 | 309.8 | 51.1 | 214,967 | 8 | 0.94 | 0.15 |
| Coliidae (5) | *Urocolius indicus* | Red-faced Mousebird | LSUMZ B34225 | 2,822,685 | 2208 | 398 | 73.9 | 877,590 | 1495 | 465.3 | 84.0 | 695,586 | 43 | 0.68 | 0.25 |
| Megalaimidae (6) | *Megalaima virens* | Great Barbet | LSUMZ B20788 | 2,302,531 | 1370 | 341 | 58.6 | 466,552 | 1174 | 351.1 | 62.7 | 412,208 | 10 | 0.86 | 0.18 |
| Picidae (7) | *Sphyrapicus varius* | Yellow-bellied Sapsucker | FLMNH 43569 | 2,693,567 | 1952 | 388 | 61.2 | 757,975 | 1542 | 416.5 | 65.9 | 642,192 | 46 | 0.79 | 0.24 |
| Phoeniculidae (8) | *Rhinopomastus* | Common Scimitarbill | LSUMZ B34262 | 1,829,285 | 1742 | 382 | 55.9 | 665,679 | 1425 | 411.1 | 59.3 | 585,753 | 24 | 0.82 | 0.32 |
| Momotidae (9) | *Momotus momota* | Blue-crowned Motmot | LSUMZ B927 | 2,694,269 | 2195 | 383 | 51.9 | 840,829 | 1587 | 430.7 | 57.3 | 682,265 | 45 | 0.72 | 0.25 |
| Trogonidae (10) | *Trogon personata* | Masked Trogon | LSUMZ B7644 | 2,371,840 | 1263 | 316 | 80.8 | 399,423 | 1117 | 315.1 | 84.6 | 351,958 | 13 | 0.88 | 0.15 |
| Tytonidae (11) | *Tyto alba* | Barn Owl | LSUMZ B19295 | 3,543,135 | 1833 | 338 | 60.7 | 620,375 | 1464 | 360.9 | 67.0 | 528,413 | 22 | 0.80 | 0.15 |
| Accipitridae (12) | *Gampsonyx swainsonii* | Pearl Kite | LSUMZ B15046 | 2,605,257 | 1588 | 525 | 64.6 | 833,617 | 1351 | 557.6 | 67.2 | 753,293 | 8 | 0.85 | 0.29 |
| Cathartidae (13) | *Cathartes aura* | Turkey Vulture | LSUMZ B17242 | 2,837,787 | 2166 | 462 | 69.4 | 1,001,122 | 1551 | 528.9 | 76.6 | 820,238 | 27 | 0.72 | 0.29 |
| Phalacrocoracidae (14) | *Phalacrocorax carbo* | Great Cormorant | LSUMZ B45740 | 4,892,448 | 1601 | 521 | 133.8 | 834,275 | 1384 | 554.1 | 137.9 | 766,906 | 10 | 0.86 | 0.16 |
| Scopidae (15) | *Scopus umbretta* | Hamerkop | LSUMZ B28330 | 3,322,061 | 2024 | 533 | 75 | 1,079,622 | 1580 | 598.1 | 78.7 | 944,999 | 46 | 0.78 | 0.28 |
| Balaenicipitidae (16) | *Balaeniceps rex* | Shoebill | LSUMZ B13372 | 1,906,136 | 1784 | 420 | 52.8 | 749,552 | 1485 | 448.9 | 55.2 | 666,057 | 19 | 0.83 | 0.35 |
| Spheniscidae (17) | *Eudyptula minor* | Little Penguin | LSUMZ B36558 | 3,009,607 | 2418 | 434 | 66.6 | 1,049,164 | 1681 | 507.5 | 73.5 | 852,753 | 42 | 0.70 | 0.28 |
| Hydrobatidae (18) | *Oceanites oceanicus* | Wilson's Storm Petrel | LSUMZ B37197 | 2,519,648 | 1930 | 488 | 73.4 | 942,397 | 1574 | 535.6 | 76.9 | 842,403 | 18 | 0.82 | 0.33 |
| Gaviidae (19) | *Gavia immer* | Common Loon | LSUMZ B7923 | 2,947,546 | 2132 | 386 | 48.4 | 821,803 | 1492 | 431.7 | 55.3 | 644,027 | 17 | 0.70 | 0.22 |
| Nyctibiidae (20) | *Nyctibius grandis* | Great Potoo | LSUMZ B15415 | 4,224,329 | 2060 | 377 | 95 | 776,650 | 1474 | 421.0 | 105.2 | 620,400 | 78 | 0.72 | 0.15 |
| Trochilidae (21) | *Colibri coruscans* | Sparkling Violetear | LSUMZ B5574 | 2,496,109 | 1881 | 384 | 64.4 | 723,418 | 1435 | 425.8 | 70.4 | 608,046 | 25 | 0.76 | 0.24 |
| Phaethontidae (22) | *Phaethon rubicauda* | Red-tailed Tropicbird | LSUMZ B35135 | 2,956,951 | 1875 | 423 | 71.2 | 792,485 | 1450 | 460.9 | 77.8 | 668,317 | 36 | 0.77 | 0.23 |
| Eurypygidae (23) | *Eurypyga helias* | Sunbittern | LSUMZ B1980 | 3,181,048 | 1988 | 416 | 78.8 | 827,124 | 1585 | 450.2 | 85.1 | 713,511 | 16 | 0.80 | 0.22 |
| Opisthocomidae (24) | *Opisthocomus hoazin* | Hoatzin | LSUMZ B9660 | 1,848,363 | 1427 | 307 | 57.9 | 438,153 | 1257 | 309.4 | 61.7 | 388,853 | 8 | 0.88 | 0.21 |
| Otididae (25) | *Ardeotis kori* | Kori Bustard | FLMNH 44254 | 2,058,864 | 2000 | 389 | 52.1 | 777,365 | 1489 | 436.0 | 57.0 | 649,136 | 54 | 0.74 | 0.32 |
| Musophagidae (26) | *Tauraco erythrolophus* | Red-crested Turaco | LSUMZ B5354 | 3,031,838 | 2134 | 402 | 70 | 858,470 | 1571 | 447.8 | 78.4 | 702,976 | 37 | 0.74 | 0.23 |
| Columbidae (27) | *Treron vernans* | Pink-necked Green Pigeon | LSUMZ B47229 | 1,949,899 | 1771 | 370 | 46.4 | 655,866 | 1337 | 409.7 | 48.5 | 547,817 | 47 | 0.75 | 0.28 |
| Pteroclididae (28) | *Pterocles exustus* | Chestnut-bellied Sandgrouse | LSUMZ B20765 | 2,167,890 | 1303 | 341 | 71.7 | 444,614 | 1130 | 351.0 | 75.5 | 396,601 | 30 | 0.87 | 0.18 |
| Phoenicopteridae (29) | *Phoenicopterus* | Chilean Flamingo | LSUMZ B37257 | 2,826,576 | 1878 | 371 | 68.4 | 696,317 | 1486 | 400.5 | 73.9 | 595,072 | 56 | 0.79 | 0.21 |
| Podicipedidae (30) | *Podiceps auritus* | Horned Grebe | LSUMZ B19296 | 2,929,983 | 1502 | 391 | 77.4 | 587,752 | 1296 | 402.1 | 79.7 | 521,175 | 2 | 0.86 | 0.18 |
| Charadriidae (31) | *Phegornis mitchelli* | Diademed Sandpiper-plover | LSUMZ B103926 | 2,488,988 | 1892 | 355 | 65.5 | 671,797 | 1518 | 381.9 | 70.3 | 579,714 | 49 | 0.80 | 0.23 |
| Psophiidae (32) | *Psophia crepitans* | Grey-winged Trumpeter | LSUMZ B7513 | 2,224,282 | 2010 | 368 | 64.9 | 739,996 | 1550 | 401.9 | 70.2 | 622,967 | 26 | 0.77 | 0.28 |

[1] Potential paralogs that were removed from the data set

[2] The number of contigs aligned to UCE loci / the total number of contigs

[3] The number of reads aligning to UCE loci / total reads

FIGURE LEGENDS

**Figure 1. Neoaves species used in this study.** Species are listed in Table 1. Numbers match those in table and on the tips of phylogenies. Illustrations are based on photos (see Acknowledgments).

**Figure 2. Relationships in Neoaves. A.** Phylogeny based on 1,541 loci from 32 species and an alignment that was 87% complete. **B.** Phylogeny based on 416 loci in 29 species and an alignment that was 100% complete. **A, B.** Branch lengths are not shown to permit easier interpretation of the topology (see Fig. 4 for phylogram of 416-locus tree and Fig. S2 for phylogram of 1,541-locus tree). Bayesian trees are shown (nodes < 0.90 PP collapsed) with circles on nodes indicating level of support for each node and congruence with the ML trees (see legend in figure). Support is shown for nodes that have less than 1.0 PP or less than 100% ML bootstrap support (PP | ML). If only a bootstrap score is shown (e.g., 46), then PP for that node = 1.0. NP = node not present in ML tree. Thus, white nodes with no values indicate 1.0 | NP.

**Figure 3. Species tree estimated from 416 individual UCE gene trees.** We collapsed nodes receiving less than 40% bootstrap support.

**Figure 4. Indels on the phylogram of the 416-locus Bayesian tree.** Hash marks indicate the phylogenetic position of the 13 indels that supported clades found in the DNA sequence data trees. The number of indels supporting each clade is shown.

SUPPORTING INFORMATION LEGENDS

**Table S1. Indels greater than 1 bp. Informative indels (n=13) that corroborate Bayesian phylogeny are indicated with bold names.**

**Figure S1. Fully resolved trees from the 1,541 locus analysis with support values. A.** Bayesian tree. **B.** Maximum-likelihood tree.

**Figure S2. Phylogram of the 1,541 locus Bayesian tree.**

**Figure S3. Fully resolved trees from the 416 locus analysis with support values. A.** Bayesian tree. **B.** Maximum-likelihood tree. **C.** Species tree.

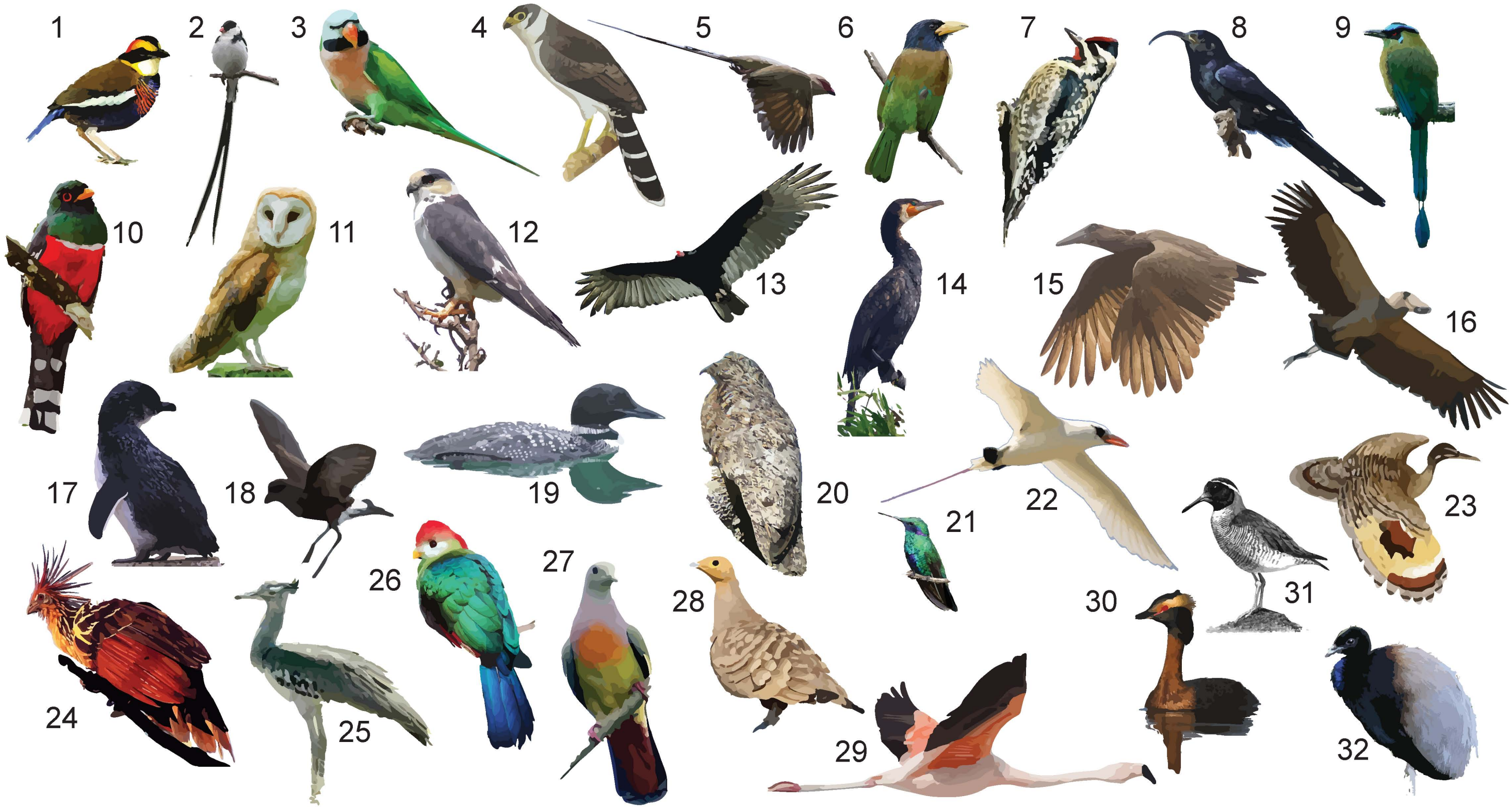

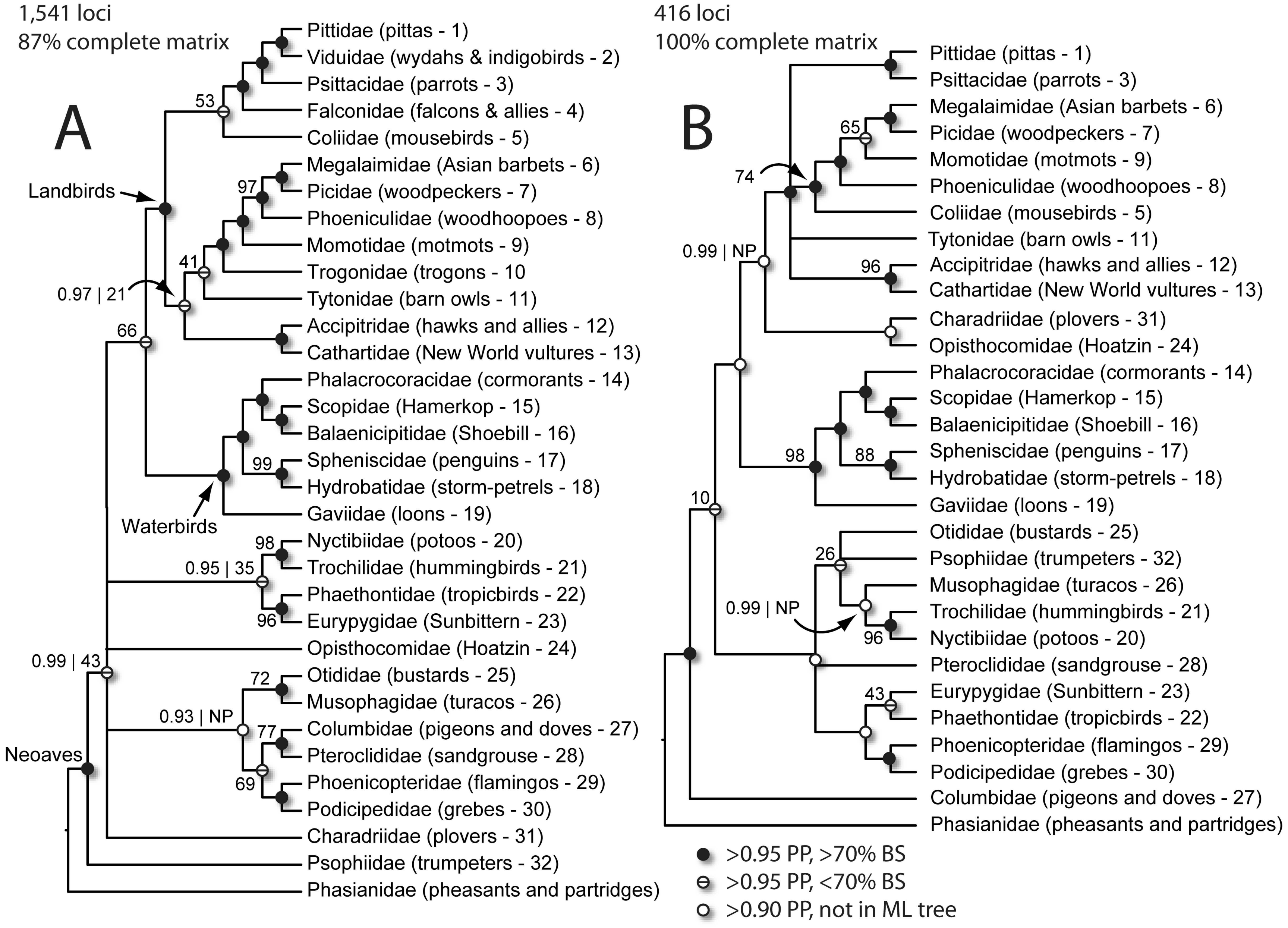

1,541 loci
87% complete matrix
A
Pittidae (pittas - 1)
Viduidae (wydahs & indigobirds - 2)
Psittacidae (parrots - 3)
Falconidae (falcons & allies - 4)
Coliidae (mousebirds - 5)
Megalaimidae (Asian barbets - 6)
Picidae (woodpeckers - 7)
Phoeniculidae (woodhoopoes - 8)
Momotidae (motmots - 9)
Trogonidae (trogons - 10
Tytonidae (barn owls - 11)
Accipitridae (hawks and allies - 12)
Cathartidae (New World vultures - 13)
Phalacrocoracidae (cormorants - 14)
Scopidae (Hamerkop - 15)
Balaenicipitidae (Shoebill - 16)
Spheniscidae (penguins - 17)
Hydrobatidae (storm-petrels - 18)
Gaviidae (loons - 19)
Nyctibiidae (potoos - 20)
Trochilidae (hummingbirds - 21)
Phaethontidae (tropicbirds - 22)
Eurypygidae (Sunbittern - 23)
Opisthocomidae (Hoatzin - 24)
Otididae (bustards - 25)
Musophagidae (turacos - 26)
Columbidae (pigeons and doves - 27)
Pteroclididae (sandgrouse - 28)
Phoenicopteridae (flamingos - 29)
Podicipedidae (grebes - 30)
Charadriidae (plovers - 31)
Psophiidae (trumpeters - 32)
Phasianidae (pheasants and partridges)
Landbirds
Waterbirds
Neoaves
53
97
41
0.97 | 21
66
99
98
0.95 | 35
96
0.99 | 43
72
0.93 | NP
77
69
416 loci
100% complete matrix
B
Pittidae (pittas - 1)
Psittacidae (parrots - 3)
Megalaimidae (Asian barbets - 6)
Picidae (woodpeckers - 7)
Momotidae (motmots - 9)
Phoeniculidae (woodhoopoes - 8)
Coliidae (mousebirds - 5)
Tytonidae (barn owls - 11)
Accipitridae (hawks and allies - 12)
Cathartidae (New World vultures - 13)
Charadriidae (plovers - 31)
Opisthocomidae (Hoatzin - 24)
Phalacrocoracidae (cormorants - 14)
Scopidae (Hamerkop - 15)
Balaenicipitidae (Shoebill - 16)
Spheniscidae (penguins - 17)
Hydrobatidae (storm-petrels - 18)
Gaviidae (loons - 19)
Otididae (bustards - 25)
Psophiidae (trumpeters - 32)
Musophagidae (turacos - 26)
Trochilidae (hummingbirds - 21)
Nyctibiidae (potoos - 20)
Pteroclididae (sandgrouse - 28)
Eurypygidae (Sunbittern - 23)
Phaethontidae (tropicbirds - 22)
Phoenicopteridae (flamingos - 29)
Podicipedidae (grebes - 30)
Columbidae (pigeons and doves - 27)
Phasianidae (pheasants and partridges)
65
74
0.99 | NP
96
98
88
10
26
0.99 | NP
96
43
>0.95 PP, >70% BS
>0.95 PP, <70% BS
>0.90 PP, not in ML tree

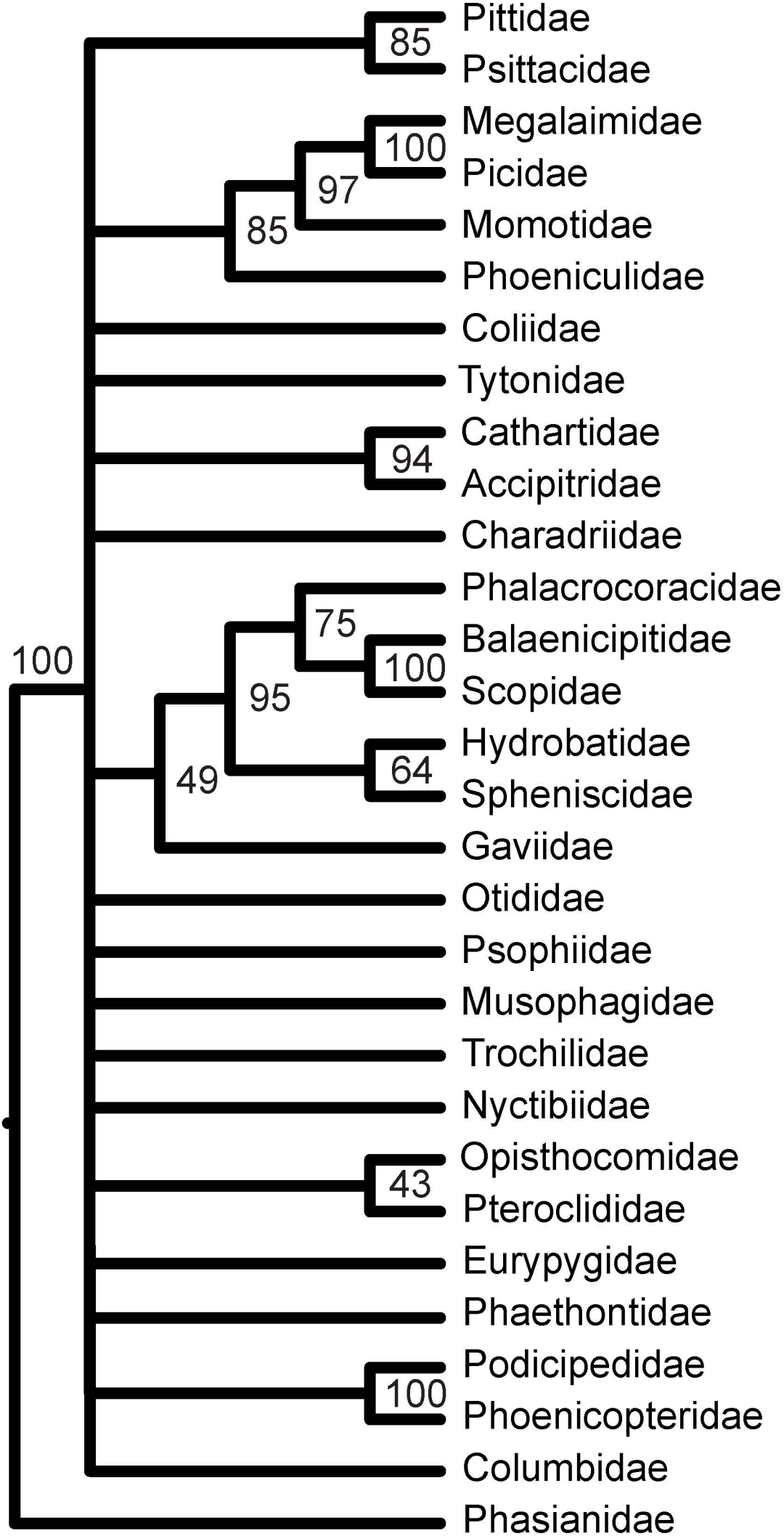

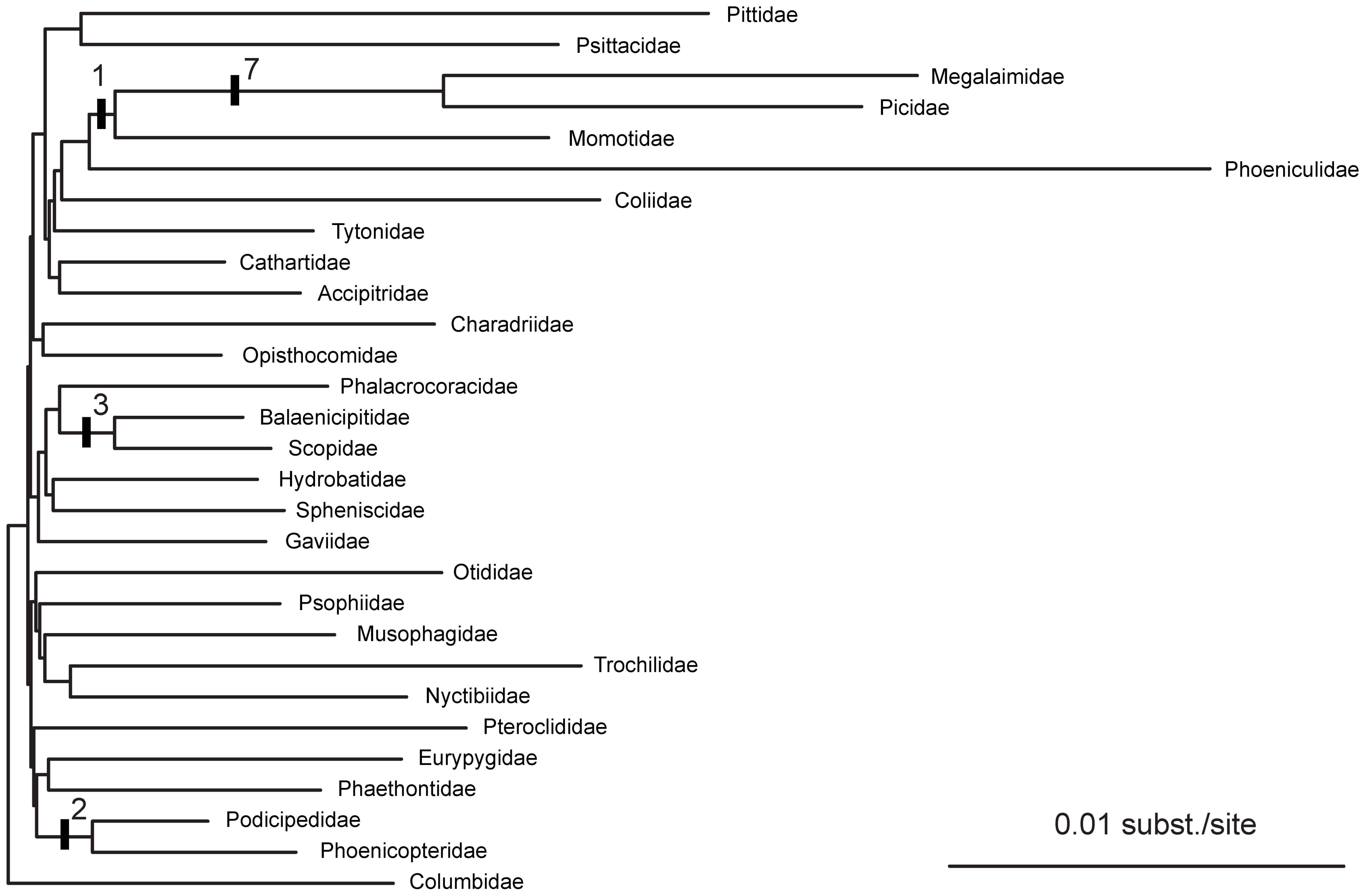

**Table S1. Indels greater than 1 bp. Informative indels (n=13) that corroborate Bayesian phylogeny are indicated with bold names.**

| UCE[1] | size[2] | type[3] | Species (informative indels in bold) |
|---|---|---|---|
| chr8_4091 | 2 | deletion | *Rhinopomastus, Sphyrapicus* |
| chr1_32309 | 3 | insertion | *Pitta, Rhinopomastus, Psittacula, Momotus, Podiceps,* |
| chr3_5661 | 2 | insertion | *Rhinopomastus, Sphyrapicus* |
| chr3_5661 | 3 | deletion | *Eurypyga, Opisthocomus* |
| chr13_707 | 6 | deletion | *Eurypyga, Treron* |
| chr9_3551 | 4 | deletion | *Colibri, Rhinopomastus, Treron, Eurypyga* |
| chr9_3551 | 7 | deletion | ***Megalaima, Sphyrapicus*** |
| chr9_3551 | 3 | deletion | *Psittacula, Ardeotis* |
| chr2_21162 | 4 | deletion | *Opisthocomus, Treron, Phoenicopterus, Podiceps* |
| chr13_2902 | 3 | insertion | *Gampsonyx, Phalacrocorax* |
| chr7_6244 | 5 | insertion | *Balaeniceps, Phalacrocorax* |
| chr2_3317 | 4 | deletion | ***Scopus, Balaeniceps*** |
| chr15_3386 | 4 | deletion | *Psittacula, Gampsonyx* |
| chr15_3386 | 4 | deletion | *Urocolius, Scopus* |
| chr1_32247 | 4 | deletion | *Momotus, Urocolius* |
| chr1_32247 | 4 | deletion | ***Phoenicopterus, Podiceps*** |
| chr3_5522 | 10 | deletion | *Sphyrapicus, Phaethon* |
| chr5_10912 | 2 | deletion | ***Megalaima, Sphyrapicus*** |
| chr2_23600 | 5 | insertion | ***Megalaima, Sphyrapicus*** |
| chr7_10289 | 2 | deletion | *Momotus, Sphyrapicus* |
| chr8_5177 | 6 | deletion | *Megalaima, Urocolius* |
| chr1_32424 | 2 | deletion | *Colibri, Ardeotis* |
| chr6_4126 | 6 | insertion | *Colibri, Pterocles, Rhinopomastus, Gampsonyx, Podiceps, Psophia* |
| chr6_4126 | 4 | insertion | *Pitta, Gampsonyx* |
| chr12_1611 | 4 | deletion | ***Momotus, Sphyrapicus, Megalaima*** |
| chr2_12990 | 4 | deletion | ***Megalaima, Sphyrapicus*** |
| chr3_19997 | 2 | deletion | *Rhinopomastus, Urocolius, Psophia* |
| chr7_10443 | 3 | deletion | *Megalaima, Treron, Sphyrapicus* |
| chr8_4221 | 3 | deletion | *Rhinopomastus, Motmotus, Sphyrapicus* |
| chr1_15632 | 3 | deletion | *Sphyrapicus, Megalaima, Opisthocomus* |
| chr11_3419 | 3 | deletion | *Balaeniceps, Motmotus, Gampsonyx* |
| chr7_10549 | 4 | deletion | *Tauraco, Phalacrocorax* |
| chr15_2007 | 2 | deletion | *Sphyrapicus, Megalaima, Psittacula, Tauraco, Podiceps* |
| chr9_3633 | 6 | deletion | ***Scopus, Balaeniceps*** |
| chr2_18663 | 2 | deletion | *Rhinopomastus, Eurypyga* |
| chr6_8088 | 4 | deletion | *Nyctibius, Psittacula, Oceanites* |
| chr1_28710 | 3 | deletion | *Sphyrapicus, Eudyptyla* |
| chr1_28710 | 3 | deletion | ***Sphyrapicus, Megalaima*** |
| chr11_4777 | 3 | deletion | ***Phoenicopterus, Podiceps*** |
| chr5_14389 | 2 | deletion | ***Megalaima, Sphyrapicus*** |
| chr1_5427 | 2 | deletion | ***Balaeniceps, Scopus*** |
| chr5_2017 | 2 | deletion | ***Megalaima, Sphyrapicus*** |
| chr2_18589 | 2 | deletion | *Cathartes, Psophia* |
| chr2_18589 | 2 | deletion | *Rhinopomastus, Psittacula, Ardeotis* |

1 Location relative to chicken genome
2 in base pairs
3 relative to chicken outgroup

**Figure S1. Fully resolved trees from the 1,541 locus analysis with support values. A.** Bayesian tree. **B.** maximum-likelihood tree.

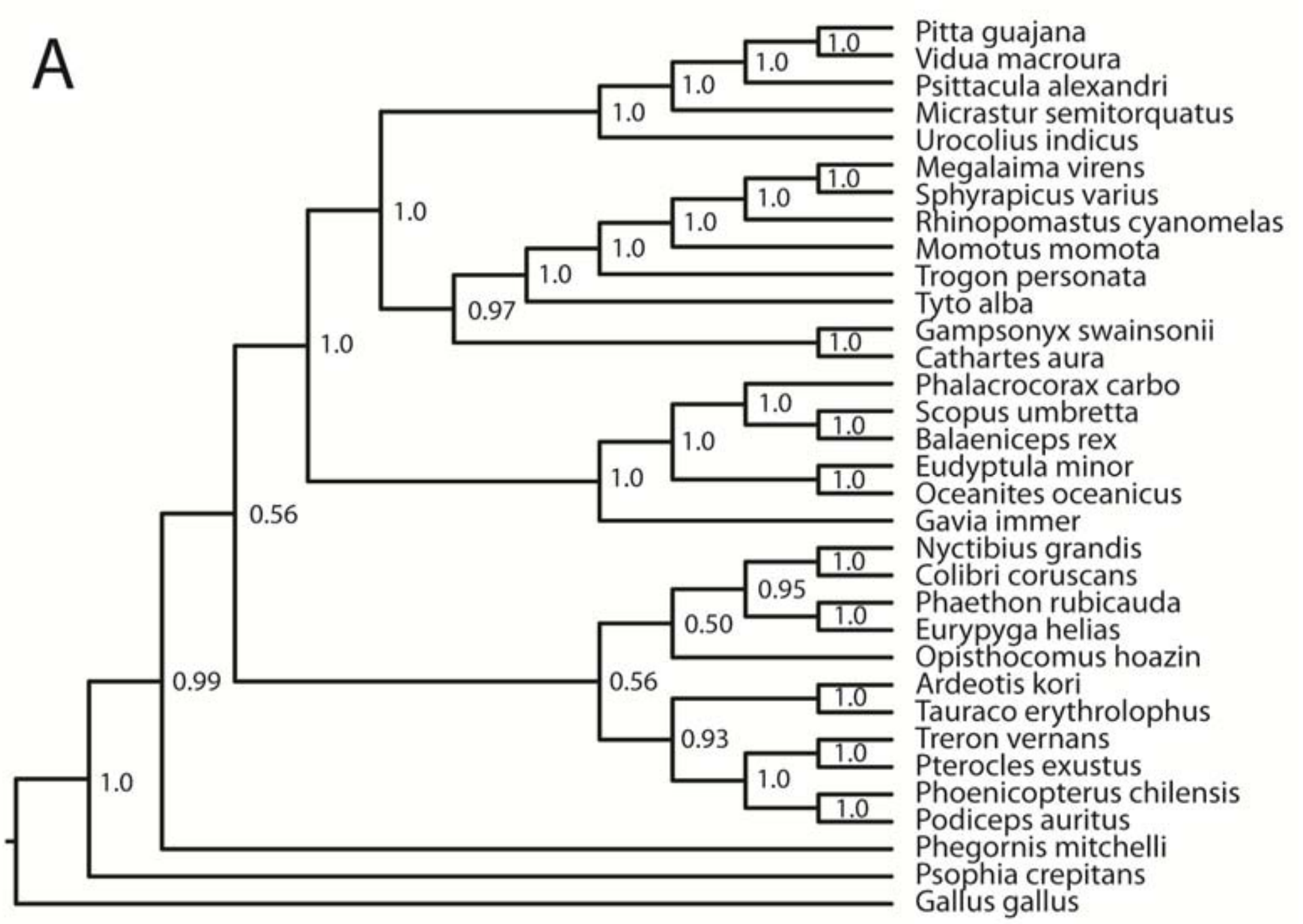

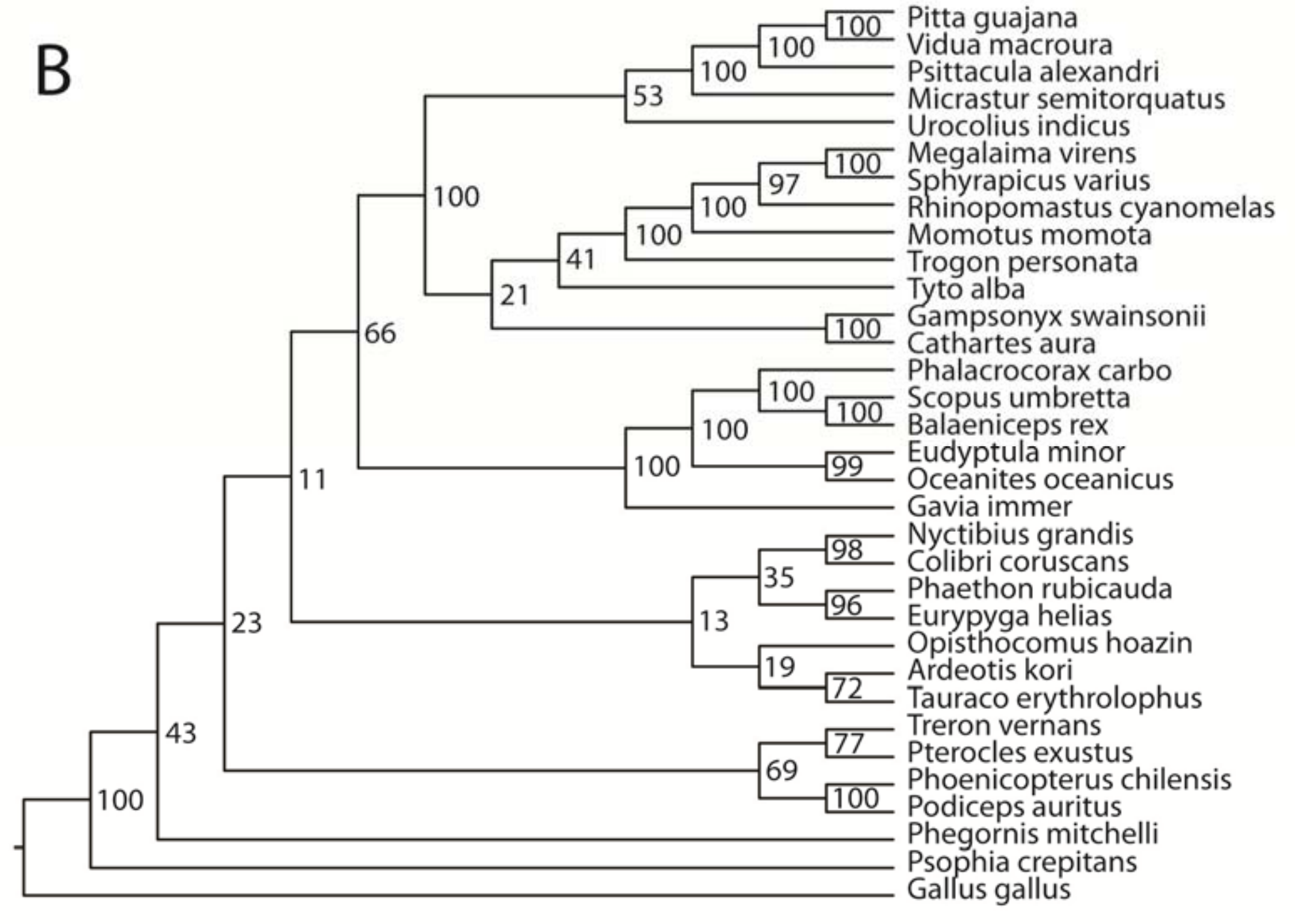

**Figure S2. Phylogram of the 1,541 locus Bayesian tree.**

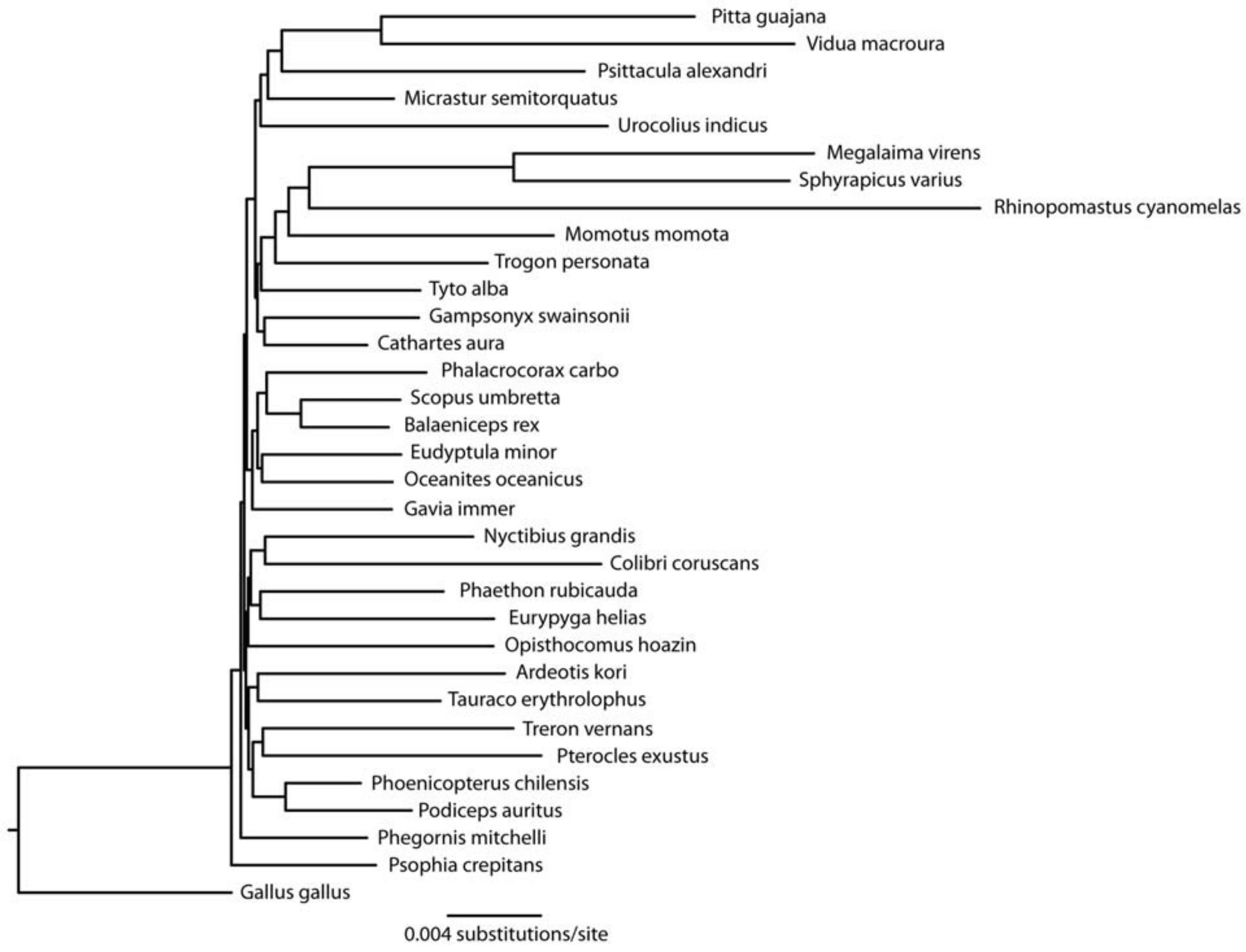

**Figure S3. Fully resolved trees from the 416 locus analysis with support values. A.** Bayesian tree. **B.** Maximum-likelihood tree. **C.** Species tree.

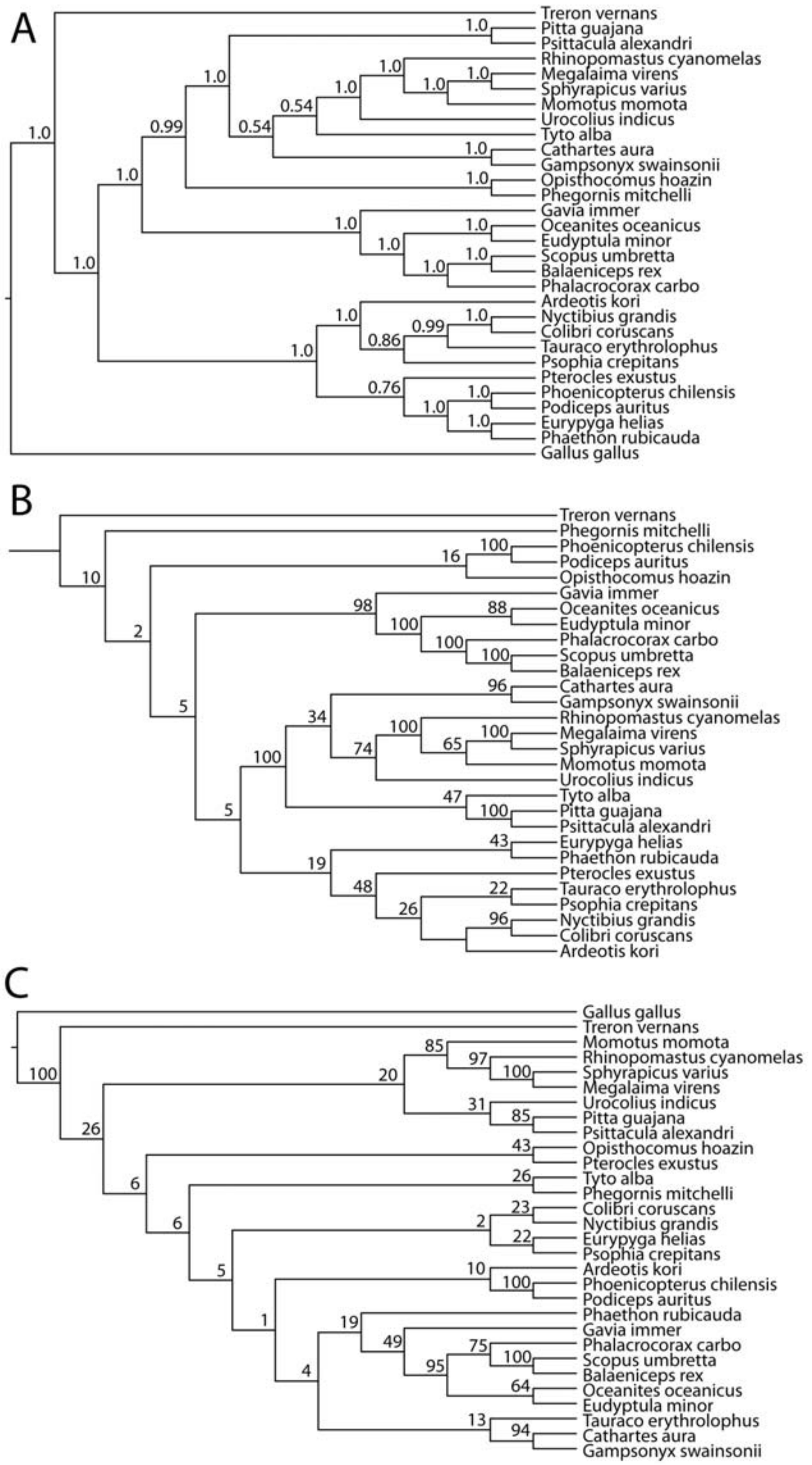